# 2D Materials in Electro-optic Modulation: energy efficiency, electrostatics, mode overlap, material transfer and integration

Zhizhen Ma[1], Rohit Hemnani[1], Ludwig Bartels[2], Ritesh Agarwal[3], Volker J. Sorger[1]

[1]Department of Electrical and Computer Engineering, George Washington University
800 22nd St., Science & Engineering Hall, Washington, DC 20052, USA
[2]Chemistry and Materials Science & Engineering, University of California, Riverside, California 92521, USA
[3]Department of Materials Science and Engineering, University of Pennsylvania,
Philadelphia, PA 19104, USA
*corresponding author, E-mail: sorger@gwu.edu

## Abstract

Here we discuss the physics of electro-optic modulators deploying 2D materials. We include a scaling laws analysis showing how energy-efficiency and speed change for three underlying cavity systems as a function of critical device length scaling. A key result is that the energy-per-bit of the modulator is proportional to the volume of the device, thus making the case for submicron-scale modulators possible deploying a plasmonic optical mode. We then show how Graphene's Pauli-blocking modulation mechanism is sensitive to the device operation temperature, whereby a reduction of the temperature enables a 10x reduction in modulator energy efficiency. Furthermore, we show how the high index tunability of Graphene is able to compensate for the small optical overlap factor of 2D-based material modulators, which is unlike classical Silicon-based dispersion devices. Lastly we demonstrate a novel method towards a 2D material printer suitable for cross-contamination free and on-demand printing. The latter paves the way to integrate 2D materials seamlessly into taped-out photonic chips.

## 1. Physics of Electro-optic Modulators

The performance of a modulator is intricately linked to the underlying physics of this opto-electronic devices [1-6], and emitters [7]. The key goal of this device is to create the highest optical index change with the least amount of voltage applied over the shortest device length. The figure of merit of modulators can therefore be stated by the data handling capacity (i.e. $f_{3dB}$ speed) divided by the energy efficiency times the device area.

2D materials are considered here as the active (i.e. modulated) materials. The stark contrasting aspect ratio of such materials lead two challenges with respect to the optical overlap factor, $\Gamma$, defined as the fraction of the optical field overlapping with the active material. This factor has two components; namely, the optical overlap factor at the wavelength of light, and the field overlap for the applied RF (i.e. modulated) field. Both need to be considered for maximum efficiency.

Next we introduce scaling laws for electro-optic modulators, asking how does the energy efficiency and speed performance change when devices are shrunk to sub micro scales.

### 1.1. Modulator Energy Efficiency

High-speed EOMs need to be power efficient and ultra-compact in order to reduce the capacitance and power consumption. Here we consider non-traveling wave modulators (i.e. lumped element devices). Given the strong light-matter-interaction enabled by the strong electric field of the plasmonic sub-diffraction limited optical mode, the effective devices lengths are on the order of micrometers as suppose to millimeters in Silicon or III-V-based modulators. The much longer wavelength of the modulated signal (typically 10's GHz), and short device length justifies utilizing the lumped element condition. The energy efficiency of an EOM is then given by the charging the device capacitor, ½ $CV^2$, where $C$ is the device capacitance, and $V$ is the driving voltage [8]. This expression takes the resistive energy loss of the modulator device itself into account during the charge-discharge cycles, ignores the power consumed by the driver circuit, and hence provides a lower bound for the switching energy [9]. For field-effect devices the refractive index change is governed by an electric field, $E$, where $E = V/h$, $h$ is the thickness of a device volume. The device capacitance is calculated by $C = \varepsilon_o \varepsilon_r wl/$. The electrical energy efficiency, $E_{Elec}$, can be further expressed by

$$E_{Elec} = \frac{1}{2}\varepsilon_o\varepsilon_r \cdot E^2 \cdot whl = \frac{1}{2}\varepsilon_o\varepsilon_r \cdot E^2 \cdot (Volume) \quad (1)$$

Taken the Pockel's effect per example, the index change for such an EOM is related to electric field by $\Delta n_p = \frac{1}{2}r_{EO}n^3 E$, $r_{EO}$ is the linear electro-optic coefficient (i.e., Pockels coefficient) of the cavity material. The resonant wavelength shift of the cavity $\Delta\lambda$ exhibits a fast response to $E-field$ by $\Delta\lambda = \frac{\Delta n_{eff}}{n_g}\cdot \lambda_R \approx \frac{1}{2}r_{EO}n^2\lambda_R E$ through taking the group index $n_g$ to be equal to the effective index [9]. The photon lifetime is related to $Q$ by $\tau_{phot} = \frac{Q \cdot \lambda_R}{2\pi c}$, and $Q$ is defined by $Q = \lambda_R/\delta\lambda$. Assuming the cavity linewidth approximated to its bandwidth, $BW$, $Q$ factor can be further expressed by $Q \approx \frac{2\pi c}{\lambda_R \cdot BW}$. We note that BW is the modulation speed in Hz. For a cavity based EOM, the condition of $\Delta\lambda > \delta\lambda$ was taking into account. We obtain the electric field,

$$E > \frac{\lambda_R \cdot BW}{\pi r_{EO} n^2 c}. \quad (2)$$

Then, the electrical energy efficiency can thus be bounded by a quality factor,

$$\begin{aligned}
Energy_{Elect.} &= \tfrac{1}{2} C V_{bias}^2 \\
&= \tfrac{1}{2}\left(\varepsilon \frac{WL}{h}\right)(h)^2 E_{critical}^2 \\
&= \tfrac{1}{2} \varepsilon \cdot E_{critical}^2 \cdot (WLh) \\
&= \tfrac{1}{2} \varepsilon \cdot E_{critical}^2 \cdot (Volume) \\
&> \tfrac{1}{2} \varepsilon \cdot \left(\frac{2}{r_{EO} n^2 Q}\right)^2 (Volume) \\
&\propto \frac{1}{F_p \cdot Q}
\end{aligned} \quad (3)$$

where $F_p$ is the Purcell factor, which is proportional to ($Q/Volume$). The results of the energy scaling show a general increase in the modulation energy per bit driven by the loss of the underlying cavity of the modulator, thus requiring a stronger index shift to achieve the desired signal extinction ratio leading to an increased voltage bias and thus higher energy (Fig. 1). Yet, deep sub-micrometer small device are possible while still preserving 100's of atto-Joule low efficiency (Fig. 1a).

We recently showed that the switching E/bit of EO modulators are able to approach that of electronic transistors of about $10^3$-$10^4$ in fundamental units of $k_B T$ (Fig. 2). For a polaritonic mode, high-index changing material, and cavity enhancement (Finesse = 10), 10's to 100 aJ/bit are possible, matching our earlier optoelectronics scaling law study [5]. Such switching energy is comparable to <10nm scaled FETs since the performance improvement over electronic links using WDM and medium fast EOM drivers (10's GHz) can approach 250x, depending on link details such as length [6].

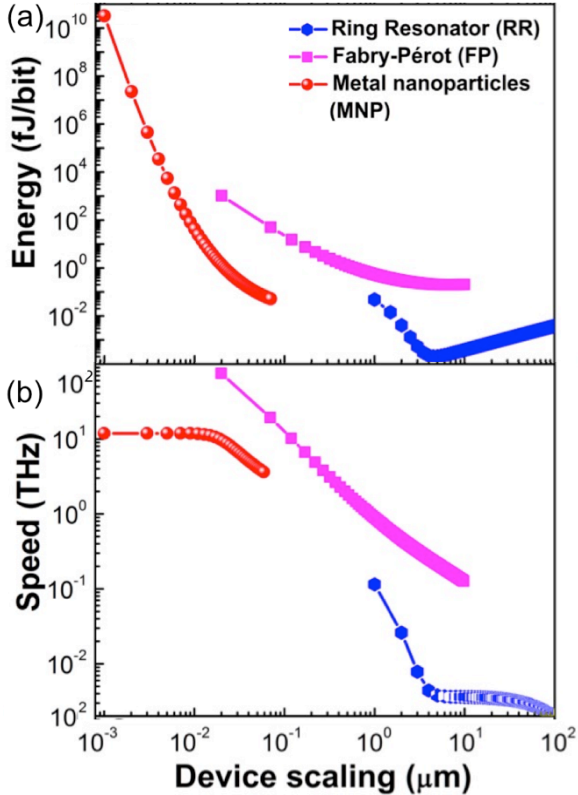

Figure 1: Electro-optic modulator scaling for three underlying device cavity systems. (a) Energy efficiency and (b) RC delay speed based on optical and electrical constrains.

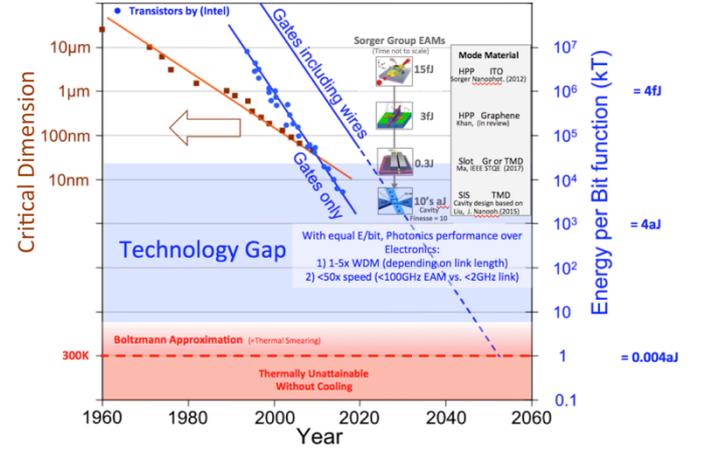

Figure 2: The other Moore's Law: Energy/bit vs. time. Optoelectronics, however, is classically limited to 10-100+ fJ/bit efficiencies, dramatically lagging electronic transistors performance. This illustratively highlights the weak light-matter-interaction. A holistic approach including material choice, optical mode, and cavity feedback enables possible sub-100aJbit EOM performance.

### 1.2. Modulator Speed

The overall modulation bandwidth of an EOM is related to the photon lifetime and $RC$-limited bandwidth (i.e., $f_{ph}$ and $f_{RC}$) through this expression [4],

$$f_{3dB} = \frac{f_{ph} f_{RC}}{\sqrt{f_{ph}^2 + f_{RC}^2}} \quad (4)$$

where $f_{ph} = \frac{1}{2\pi\tau_{phot}} \approx \frac{c}{\lambda_R Q}$, $f_{RC} = \frac{1}{2\pi(R_s + R_{dr})C_J}$, $R_s$ is the modulator series resistance, $R_{dr}$ is the driver impedance, and $C_J$ is the modulator junction capacitance, here $C_J = \varepsilon_0 \varepsilon_r \frac{wl}{h}$. Eqn. 4 indicates that the modulation bandwidth is limited by the $Q$ factor. For the comparison of EOM energy efficiency and modulation speed, we configure an EOM with cavity enhanced by microring (RR), Fabry Perot (FP), and plasmonic metal nanoparticle (MNP) cavity, respectively. The corresponding bandwidth expression of Eqn. 4 only uses each cavity's $Q$ formula for the device scaling. Our results show that ultra-fast modulation beyond 100 GHz should be possible, even when a resistive scaling model borrowed from transistors is applied. This suggests that there is much potential beyond the current 10's of GHz fast devices.

## 2. 2D Materials for electro-optic Modulation

There are a variety of 2D materials, but for electro-optic modulation Graphene's Pauli Blocking has shown decent functionality as demonstrated by. Here we also consider graphene, however other 2D materials are similarly synergistic to the design rules applied here, i.e. in-plane field components of the optical mode. Indeed efforts have been made in integrating graphene with plasmonics with the purpose of modulation [1-4,6]. Yet, the anisotropy of 2D films introduce challenges with respect to polarization alignment using plasmonics since field lines are always perpendicular to a metal plane. As a result, plasmonic approaches, thus far, have shown low modulation capability and non-synergistic device designs despite Graphene's strong index modulation potential. Nevertheless, graphene phase modulation shows tens of GHz fast modulation, however relies on the strong feedback from a mirroring cavity leading to non-compact footprints and temperature sensitivities. Thus, in this study we focus on engineering the optical mode profile of graphene to enhance the light-matter interaction while using a plasmonic modulation platform to decrease the device footprint. Previously we showed that volumetric device shrinking (scaling) of EO modulators fundamentally enables to reduce the energy consumption (E/bit), and sub fJ/bit are possible for plasmonic modulator design [5].

### 2.1. Case: Graphene and Temperature Sensitivity

Graphene is an anisotropic material given its dimensions: in its honeycomb like lattice plane, the in-plane permittivity ($\varepsilon_\parallel$) can be tuned by varying its chemical potential $\mu_c$, whereas the out-of-plane permittivity is reported to remain constant around 2.5 [9]. We model graphene with two different temperatures by Kubo model at T = 0K and T = 300K (Fig. 3). We note, that the drastic change in Graphene's imaginary refractive index, $\kappa$, is due to the strong effect through Pauli blocking, thus making graphene naturally suitable material for electro-absorption modulators. At higher temperature, the imaginary refractive index vs. chemical potential is smeared due to the natural temperature dependency of the Fermi-Dirac distribution function, leading to a sharp transition upon cooling. Doping (i.e. biasing) graphene to a chemical potential near half of the photon's energy, a small switching energy is needed for of only >30meV for T = 0K versus ~100meV at T = 300K (Fig. 3). This difference in the minimum voltage of about 3x is equivalents to an energy saving of about 10-fold improvement of devices are operated at cry-temperatures. However, the voltages required in devices for actual devices is modulated from zero-chemical potential to the Pauli-blocking regime of about $\Delta\mu_c$ = 0.4-0.5eV requiring voltage changes on the order of 5-10 Volts depending on the electrostatics and contact resistance. The latter introduces a voltage drop across the contacts, lowering the actual applied voltage range across the device capacitor. The metal contact from a plasmonic modulator offers here a unique advantage over photonic devices, since the drive voltage suffers no degradation in the contacts.

### 2.2. Optical Overlap Factor

The performance of a modulator depends linearity on the ability for the optical mode to overlap with the active material [6]. A comparison of the modulation performance as a function of modal overlap shows that the weak plasma dispersive index tuning in silicon requires almost unity high modal overlaps to achieve even only medium-high modulation. The carrier tunability of transparent-conductive-oxides (TCO) is able to deliver unity-strong index modulation, while Graphene is able to surpass this even further despite the vanishing cross-sectional overlap of Graphene with an optical mode (Fig. 4). Thus, if the optical mode is even further confined using metal optics or meta-material approaches delivering near or even sub-diffraction limited small optical mode areas, then a rather high modulation performance is expected. For squeezed modes in metallic slot-waveguides, the maximum overlap factor can be about 0.5% for 2D materials, whilst reaching 40% when a regular photonic waveguide would be filled with a combination of multiple 2D material-oxide layers (insets, Fig 4).

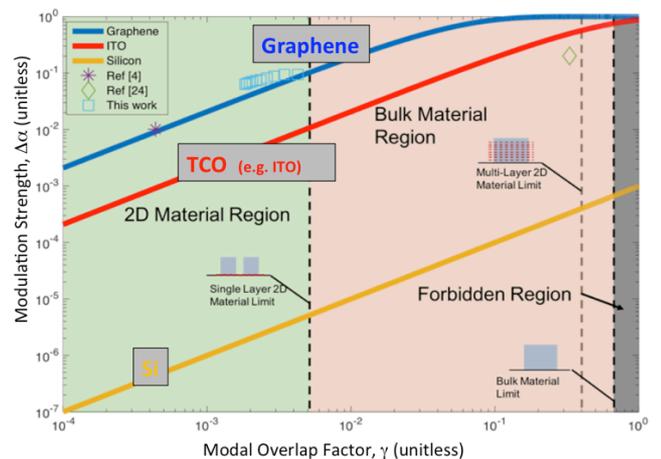

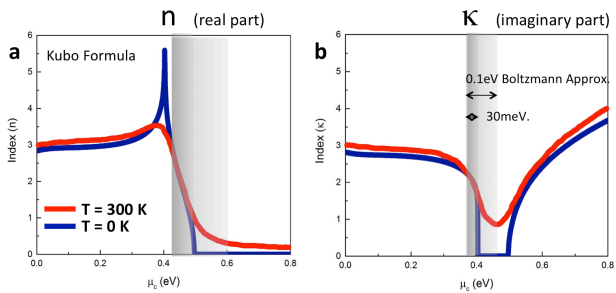

Figure 3: Refractive index for Graphene vs. chemical potential for room- and cryogenic temperatures based on the Kubo model. Gray shaded regions denote chemical potential tuning for modulation. a) Real part, phase, and (b) Imaginary part, loss. $\lambda_{photon}$ = 1550 nm (0.8eV).

Figure 4: Electro-absorption modulator performance as a function of optical mode overlap for three different active materials. The analysis shows that the high index tunability of Graphene is able to compensate the low modal overlap given Graphene's miniscule thickness.

In regard to modulation mechanisms beyond graphene, such as for TMDs or other 2D materials, this requires a detailed discussion is goes beyond this publication. However, similar to Graphene the low-dimensionality of the material and subsequent lack of Coulomb screening anticipates a high exciton modulation in these materials leading to strong index modulation.

### 3. Towards a 2D Material Printer

A key challenge at present is the ability to accurately place and integrate 2D materials into photonic integrated circuits. In detail these are (1) synergistic growth methods, (2) deterministic positioning and placement, and (3) cross-contamination-free transfer methods. Regarding the former, while CVD synthesis methods for 2D materials on-chip do exist they do have substrate limitations (i.e. $SiO_2$) and often require a high thermal budget (i.e. 600-800C) [10,11]. This is why the rather simplistic method of exfoliation is still a viable option to demonstrate functional opto-electronic devices, despite up-scaling limitations. All methods for exfoliation use some form of a sticky tape to peal off sheets of 2D materials. They then differ in the method of how these 2D materials are placed on the target substrate (Fig. 5).

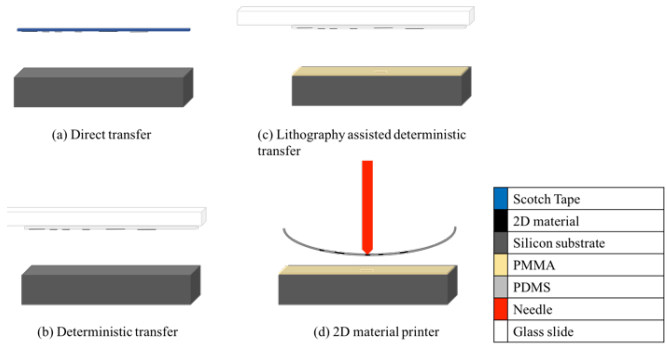

Figure 5: Schematic of four different 2D material transfer techniques utilizing exfoliation. (a) Direct transfer from scotch tape (b) Deterministic transfer technique (c) lithography assisted deterministic transfer technique (d) 2D material printer.

In the direct transfer method the 2D material residing on the exfoliated tape is directly pressed onto the silicon substrate (Fig. 5a). However, the flatness and uncontrollability of the exfoliation processes leads to transfers of unwanted flakes onto the target substrate and with no control on the position of the desired flake (Fig. 6a,b). If the latter was for example a Silicon photonics tape-out chip, then the direct transfer method has two risks; (i) mechanical scratching of waveguides and devices on-chip, and (ii) cross-contamination of the chip with other 2D materials. An improvement from this is the deterministic transfer method, where the exfoliated flake is first transferred onto a thin PDMS film, which is thence placed onto a glass cover slide, connected to a micromanipulator (Fig. 5b). Given its transparency a 2D flake can be selected and pressed (mechanical force) onto the desired location on the chip, thus improving placement accuracy (Fig. 6b). Hence, solving the accuracy issue, but not the contamination problem since the applied force is spatially large such as applied via a thumb or tweezers is the shortcoming of this method and the state of the art (Fig. 6c,d) [12].

Here we introduce to improvements with respect to transfer accuracy with minimal cross-contamination to neighboring devices or structures on the photonic chip. In the first innovation we use lithographic patterning to create openings in a photoresist into which to transfer a 2D material (Fig. 5c). That is, after exposure and development of the resist, 2D flakes transferred even without spatial alignment lead to an accurate transfer, since the unwanted transferred flakes will be washed away during the following lift-off step (Fig. 6e,f). The second innovation replaces the spatially inaccurate method of applying the transfer force with the bare hand by utilizing a fine needle making use of the stretch ability of the PDMS film (Fig. 5d). Here only at the location of the lithographic opening the PDMS film is locally pressed to transfer the previously selected 2D material flake.

We analyze the transfer cross-contamination quality of the four aforementioned transfer methods to quantify their respective improvement potentiation from one another (Fig. 6). For this we followed a statistical approach of creating the sum of the area of all unwanted flakes in a certain area around a targeted (selected) flake. Since the target flake size is somewhat random leading to flake distributions, we normalized the selection area (the area from where to count the flake cross-contamination) from which to include unwanted flakes by the area of the target flake. This enables a proper comparison across varying target flake sizes. Our results across these four methods show a clear decrease in the mean value of the contamination ratio when the lithography method is used, and another improvement when the needle is used in conjunction with the lithographic placement when compared to both the direct transfer and deterministic transfer methods (Fig. 6). Since the lithography-and-pin-assisted transfer allows for a deterministic to-be-transferred flake selection via the optically transparent PDMS film, targeted (i.e. pre-selected via optical image identification) flake transfer via the pin-pressing transfer, and overall a cross-contamination free approach, it resembles the functionality of a printer for 2D materials once automated.

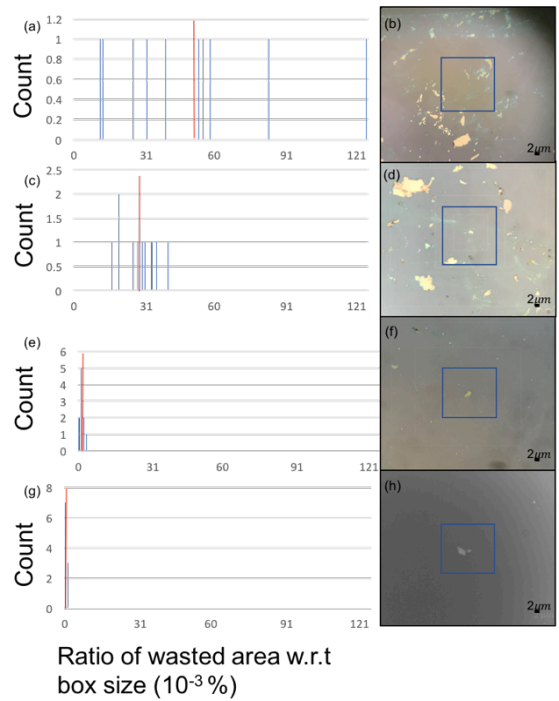

Figure 6: Statistical Analysis and comparison of the four 2D material transfer techniques from Fig. 5. (a)-(b) direct transfer using scotch tape, (c)-(d) deterministic transfer using PDMS and stiff glass slide (e)-(f) lithography assisted deterministic transfer (g)-(h) 2D material printer, Red line = mean of each data set.

Next, we applied to improved lithography-assisted pin-based method to taped-out Silicon photonics ring-resonators (Fig. 7). Comparing the cleanliness of the taped-out chip (Fig 7a) post direct-transfer method 2D material transfer, results in a dirty chip with many flakes crossing waveguides or other undesired regions such as neighboring rings etc. This is significantly improved when the lithography-and-pin assisted transfer is deployed (Fig. 7c).

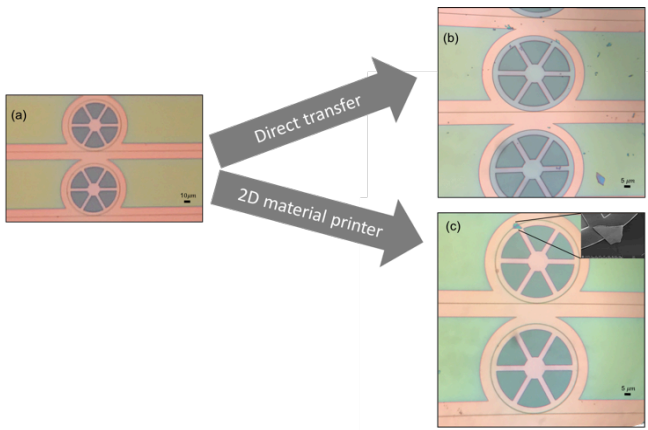

Figure 7: Cross-contamination comparison of a (a) clean taped-out Silicon photonics chip after the (b) direct-transfer using scotch tape method against the (c) 2D material printer demonstrating a chip-save non-cross-contamination clean transfer approach.

In conclusion, we have shown how the performance defined by the energy per bit and speed of electro-optic modulators can be improved when the physical volume of the device is scaled down aggressively beyond the 1-micrometer range. We then overlaid our previously demonstrated modulator energy efficiencies with transistor power consumption and showed that nanoplasmonic modulators based on 2D materials are able to approach single field-effect-transistor energy efficiency levels. This then indicates the performance improvement of photonics over electronics since the former allows for 10's of GHz fast modulation, since only the opto-electronic device needs to be charged and not the entire electrical wire, and optics allows for wavelength-division-multiplexing enabling high data parallelism [8]. Furthermore, we show how the high optical index modulation of 2D materials is able to overcompensate for the low optical overlap factor. However, for a single 2D film integrated the overlap factor approaches 1%, and 40% when an entire photonic-size mode is filled with 2D material-oxide heterostructures. Lastly, we demonstrated a path towards a deterministic, and cross-contamination free transfer method for 2D materials, resembling printer functionality. This enables higher levels of 2D integration into on-chip photonic waveguiding systems [13] by way of hybrid co-integration [14-18] of passive silicon photonics with active 2D material functionality [19,20].

## Acknowledgements

V.S., L.B are supported by the National Science Foundation under award number NSF DMREF 1436330. V.S. and R.A. are supported by the Army Research Office under contract number W911NF-16-2-0194.

## References

1. S. K. Pickus, S. Khan, C. Ye, Z. Li, and V. J. Sorger, "Silicon Plasmon Modulators: Breaking Photonic Limits" IEEE Photonic Society 27, 6 (2013).
2. K. Liu, Z.R. Li, S. Khan, C. Ye, V. J. Sorger, "Ultra-fast electro-optic modulators for high-density photonic integration" Laser & Photonics Review, 10, 11-15 (2015).
3. R. Amin, C. Suer, Z. Ma, J. Khurgin, R. Agarwal, V. J. Sorger, "Active Material, Optical Mode and Cavity Impact on electro-optic Modulation Performance" arXiv:1612.02494 (2016)
4. R. Amin, C. Suer, Z. Ma, J. Khurgin, R. Agarwal, V. J. Sorger, "A Deterministic Guide for Material and Mode Dependence of On-Chip Electro-Optic Modulator Performance", Solid-State Electronics, Special Issue, DOI: 10.1016/j.sse.2017.06.024 (2017).
5. K. Liu, A. Majumdar, V. J. Sorger, "Fundamental Scaling Laws in Nanophotonics" Scientific Reports, 6, 37419 (2016).
6. Z. Ma, M. H. Tahersima, S. Khan, V. J. Sorger, "Two-Dimensional Material-Based Mode Confinement Engineering in Electro-Optic Modulators," *IEEE Journal of Selected Topics in Quantum Electronics*, vol. 23, no. 1, 1-8 (2017).
7. M. H. Tahersima, et al. "Testbeds for Transition Metal Dichalcogenide Photonics: Efficacy of Light Emission Enhancement in Monomer vs. Dimer Nanoscale Antennas", ACS Photonics, 4, 1713-1721 (2017).

8. Miller, D.A.B. Energy consumption in optical modulators for interconnects. Opt. Express 20, A293-A308 (2012).
9. Lin, H.T., Ogbuu, O., Liu, J., Zhang, L., Michel, J. & Hu, J.J. Breaking the energy-bandwidth limit of electrooptic modulators: theory and a device proposal. J. Lightwave Technol. 31, 4029-4036 (2013).
10. J. Martinez et al. "Transport Properties of CVD Grown TMDs on Flat and Patterned Substrates", APS March Meeting Abstracts (2015).
11. J. Mann et al. "Facile growth of monolayer $MoS_2$ film areas on $SiO_2$", The European Physical Journal B, 86, 5, 226 (2013).
12. Castellanos-Gomez, A., Buscema, M., Molenaar, R., Singh, V., Janssen, L., van der Zant, H. S., Steele, G. A. "Deterministic transfer of two-dimensional materials by all-dry viscoelastic stamping", 2D Materials, 1(1), 011002 (2014).
13. S. Sun, et al. "The case for hybrid photonic plasmonic interconnects (HyPPIs): Low-latency energy-and-area- efficient on-chip interconnects", IEEE Photonics *J*. **7**, 1-14 (2015).
14. N. Li, et al., "Nano III-V Plasmonic Light-Sources for Monolithic Integration on Silicon", Nature: Scientific Reports, 5, 14067 (2015).
15. K. Liu, V. J. Sorger, "An electrically-driven Carbon nanotube-based plasmonic laser on Silicon" Optical Materials Express, 5, 1910-1919 (2015).
16. K. Liu, N. Li, D. K. Sadana, V. J. Sorger, "Integrated nano-cavity plasmon light-sources for on-chip optical interconnects" ACS Photonics, 3, 233-242 (2016).
17. A. Fratalocchi, et al., "Nano-optics gets practical: Plasmon Modulators", Nature Nanotechnology, 10, 11-15 (2015).
18. Z. Ma, Z. Li, K. Liu, C. Ye, V. J. Sorger, "Indium-Tin-Oxide for High-performance Electro-optic Modulation", Nanophotonics, 4, 1 (2015).
19. J. K. George, V. J. Sorger, "Graphene-based Solitons for Spatial Division Multiplexed Switching", Optics Letters 42, 4, 787-790 (2017).
20. B. Lee, W. Liu, C. H. Naylor, J. Park, S. Malek, J. Berger, A.T. C. Johnson, R. Agarwal, "Electrical tuning of exciton-plasmon polariton coupling in monolayer $MoS_2$ integrated with plasmonic nanoantenna lattice", Nanoletters, 17, 4541-4547 (2017).